\documentclass[a4paper,11pt]{article}
\usepackage[dvipdfmx]{graphicx}
\usepackage{amsmath}        
\usepackage{amssymb}        
\usepackage{cite}
\usepackage{cancel}
\usepackage{fancybox}
\usepackage[dvips]{color}
\usepackage{url}
\usepackage{ulem}
\usepackage{pifont}
\usepackage{lscape}
\usepackage[all,knot]{xy}
\usepackage{longtable}

\newsavebox{\BCmatrixA}
\savebox{\BCmatrixA}{$\left(\begin{array}{cc}\eta_2&\eta_3\\ \eta_0&\eta_1\end{array}\right)$}
\newsavebox{\BCmatrixB}
\savebox{\BCmatrixB}{$\left(\begin{array}{c}\eta_2\\ \eta_0\end{array}\right)$}
\newsavebox{\BCmatrix}
\savebox{\BCmatrix}{$\left(\begin{array}{cc}\eta_{2{\bf 171}\pm}^{(\prime)}\overline{\gamma}P_{2{\bf 171}\pm}^{(\prime)}&\eta_{3{\bf 171}\pm}^{(\prime)}\overline{\gamma}P_{3{\bf 171}\pm}^{(\prime)}\\ \eta_{0{\bf 171}\pm}^{(\prime)}\overline{\gamma}P_{0{\bf 171}\pm}^{(\prime)}&\eta_{1{\bf 171}\pm}^{(\prime)}\overline{\gamma}P_{1{\bf 171}\pm}^{(\prime)}\end{array}\right)$}

\usepackage{multicol}
\usepackage{colortbl}
\usepackage{multicol}

\makeatletter
 
 \@addtoreset{equation}{section}
\makeatother

\begin{document}

\title{Family Unification in Special Grand Unification}

\author{Naoki Yamatsu
\footnote{Electronic address: yamatsu@high.hokudai.ac.jp}
\\
{\it\small Institute for the Advancement of Higher Education,
Hokkaido University}\\
{\it\small Sapporo, Hokkaido 060-0817, Japan}
}
\date{\today}

\maketitle

\begin{abstract}
We discuss family unification in grand unified theory (GUT) based on an
$SU(19)$ GUT gauge group broken to its subgroups including a special
subgroup.
In the $SU(19)$ GUT on the six-dimensional (6D) orbifold space
$M^4\times T^2/\mathbb{Z}_2$, 
three generations of the 4D SM Weyl fermions can be embedded into a 6D
bulk Weyl fermions in an $SU(19)$ second-rank anti-symmetric tensor
representation. 6D and 4D gauge anomalies can be canceled out by
considering proper matter content without 4D exotic chiral fermions at
low energies.
\end{abstract}

\section{Introduction}

The existence of three chiral generations of quarks and leptons is one
of the most mysterious fact in particle physics.
In addition, their hierarchical mass structures generated by
the Higgs mechanism via their corresponding Yukawa couplings strongly
suggest the existence of a hidden structure of nature.
There are many attempts to understand the origin of chiral generations
and/or their hierarchical masses structures by considering e.g.,
so-called horizontal symmetry (or family symmetry) in four-dimensional
(4D) theories 
\cite{Wilczek:1978xi,Froggatt:1978nt,Yanagida:1979as,Maehara:1979kf,Inoue:1994qz,King:2001uz,Maekawa2004},
geometrical structures in higher-dimensional theories
\cite{Yoshioka:1999ds,Fujimoto:2012wv,Fujimoto:2017lln}
and string theories
\cite{Abe:2008sx,Abe:2015yva,Mizoguchi:2014gva}.

As is well-known, quarks and leptons for each generation in the Standard
Model (SM) can be unified into one multiplet (or two multiplets) in
grand unified theories (GUTs) \cite{Georgi:1974sy}.
There are many GUTs in 4D framework
\cite{Georgi:1974sy,Inoue:1977qd,Fritzsch:1974nn,Ida:1980ea,Fujimoto:1981bv,Gursey:1975ki}
and higher-dimensional space framework
\cite{Kojima:2011ad,Kojima:2016fvv,Burdman:2002se,Lim:2007jv,Kim:2002im,Fukuyama:2008pw,Hosotani:2015hoa,Hosotani:2015wmb,Yamatsu:2015rge,Furui:2016owe,Hosotani:2016njs,Hosotani:2017ghg,Hosotani:2017edv}
(For review, see Refs.~\cite{Slansky:1981yr,Yamatsu:2015gut}.)

There are some attempts to unify GUT and family groups into a larger GUT
group in 4D and higher-dimensional theories 
\cite{Ramond:1979py,Kawamura:2007cm,Kawamura:2009gr,Goto:2013jma,Albright:2016lpi,Goto:2017zsx,Reig:2017nrz,Reig:2018ocz}.
However, such attempts are based on GUT groups and their limited
subgroups so-called {\it regular subgroups}: e.g.,
$E_8\supset E_7\supset E_6\supset SO(10)\supset SU(5)\supset
G_{\rm SM}(:=SU(3)_C\times SU(2)_L\times U(1)_Y)$.
There are other subgroups called {\it special subgroups}
(or {\it non-regular subgroups}): e.g.,
$SO(248)\supset E_8$, $USp(56)\supset E_7$, $SU(27)\supset E_6$,
and $SU(16)\supset SO(10)$.
(For Lie groups and their subgroups, see e.g.,
Refs.~\cite{Dynkin:1957ek,Dynkin:1957um,Mckay:1977,McKay:1981,Slansky:1981yr,Cahn:1985wk,Fonseca:2011sy,Feger:2012bs,Yamatsu:2015gut}.)

Recently, new-type GUTs called {\it special GUT} based on GUT groups
$SO(32)$ and $SU(16)$ and their special subgroup $SO(10)$ have been
proposed in Refs.~\cite{Yamatsu:2017sgu,Yamatsu:2017ssg}.
The main results of $SO(32)$ and $SU(16)$ special GUTs are summarized as
below.
In an $SU(16)$ special GUT based on its GUT group $SU(16)$ broken
to its special subgroup $SO(10)$, a 4D $SU(16)$ ${\bf 16}$ Weyl fermion
can be identified with one generation of quarks and leptons; 4D $SU(16)$
gauge anomaly cancellation does not work in 4D framework, while it works
in 6D framework without any exotic chiral fermions once we take into
account $SU(16)$ symmetry breaking effects \cite{Yamatsu:2017sgu}.
Almost the same results are obtained in an $SO(32)$ special GUT
\cite{Yamatsu:2017ssg}. 

In special GUT framework, family unification can be considered by using 
GUT groups and their ``regular-type'' and ``product-type'' subgroups;
the former example is
$SU(19)\supset SU(16)\times SU(3)\times U(1)$;
the latter example is 
$SU(48)\supset SU(16)\times SU(3)$,
where $SU(16)$ contains an ordinary GUT gauge group $SO(10)$ and $SU(3)$
is a family gauge group. 
(Their branching rules of
$SU(19)\supset SU(16)\times SU(3)\times U(1)$,
$SU(48)\supset SU(16)\times SU(3)$,
etc. can be calculated e.g., by
using a projection matrix method 
shown in Refs.~\cite{Mckay:1977,McKay:1981,Yamatsu:2015gut}.)

First, for a regular-type case,
an example of GUT gauge groups and their subgroup pair is
$SU(19)\supset SU(16)\times SU(3)\times U(1)
\supset SO(10)\times SU(3)\times U(1)$.
It is a simple extension of $SU(16)$.
The branching rule of
$SU(19)\supset SU(16)\times SU(3)\times U(1)
\supset SO(10)\times SU(3)\times U(1)$
for the $SU(19)$ defining representation is 
\begin{align}
{\bf 19}=&
({\bf 16,1})(3)
\oplus({\bf 1,3})(-16).
\label{branching-rule-19}
\end{align}
In this case, e.g., an $SU(19)$ second-rank anti-symmetric tensor
representation contains three generations of quarks and leptons.
Its branching rule is given by
\begin{align}
{\bf 171}=&
({\bf 16,3})(-13)
\oplus({\bf 120,1})(6)
\oplus({\bf 1,\overline{3}})(-32),
\label{branching-rule-171}
\end{align}
where an $SU(16)$ ${\bf 120}$ representation is complex while
an $SO(10)$ ${\bf 120}$ representation is real.
A 4D Weyl fermion in an $SO(10)$ ${\bf 120}$ representation is
vectorlike, so when we take into account symmetry breaking
effects for $SU(19)$ to $SO(10)$, only three 4D $SO(10)$ ${\bf 16}$ Weyl
fermions remain chiral.
Also, an $SU(19)$ second-rank symmetric tensor representation ${\bf 190}$
contains an $SU(16)\times SU(3)$ $({\bf 16,3})$ representation.
Note that the $SU(19)$ ${\bf 190}$ contains unwilling
$SU(16)\times SU(3)$ complex representations.
The $SU(19)$ adjoint representation ${\bf 360}$
contains not only  an $SU(16)\times SU(3)$
$({\bf 16,\overline{3}})$ but also its conjugate representation
$({\bf \overline{16},{3}})$.
(In Ref.~\cite{Fonseca:2015aoa}, R.M.~Fonseca has also pointed out that
a 4D Weyl fermion in an $SU(19)$ ${\bf 171}$ representation contains the
SM fermions plus vectorlike particles only, which was found by using the
Susyno program \cite{Fonseca:2011sy}.)

Next, for a product-type case, an example of GUT gauge groups and their
subgroup pair is
$SU(48)\supset SU(16)\times SU(3)\supset SO(10)\times SU(3)$.
The branching rule for the $SU(48)$ defining representation is 
\begin{align}
{\bf 48}=&({\bf 16,3}).
\end{align}
The 4D Weyl fermion in the $SU(48)$ defining representation can be
identified with three chiral generations of quarks and leptons.
As the best of my knowledge, there is no way to construct an $SU(48)$
gauge theory that contains only three chiral generations
without any gauge anomalies at least for 4D, 5D, 6D framework.
We will not discuss its possibility in this paper. 

In this paper, we discuss a 6D $SU(19)$ special GUT on
$M^4\times T^2/\mathbb{Z}_2$, whose GUT group includes an $SO(10)$ GUT
group and an $SU(3)_F$ family group. The main purpose of this paper is
to show that three generations of the 4D SM Weyl fermions can be
embedded into a 6D bulk Weyl fermions in an $SU(19)$ second-rank
anti-symmetric tensor representation.
6D and 4D gauge anomalies can be canceled out by considering
proper matter content without 4D exotic chiral fermions at low energies.

This paper is organized as follows. In Sec.~\ref{Sec:basics}, before we
discuss a special GUT based on an $SU(19)$ gauge group, we discuss basic
properties of $SU(19)$ and its subgroups mainly by using technique in
Ref.~\cite{Yamatsu:2015gut}.
In Sec.~\ref{Sec:Special-GUT}, 
we construct a 6D $SU(19)$ special GUT on $M^4\times T^2/\mathbb{Z}_2$.
Section~\ref{Sec:Summary-discussion} is devoted to a summary and
discussion.

\section{Basics for $SU(19)$ and its subgroups}
\label{Sec:basics}

First, we check how to embed three generations of the SM Weyl fermions
into 4D Weyl fermion in an $SU(19)$ second-rank anti-symmetric tensor
representation ${\bf 171}$.
For regular and special embeddings
$SU(19)\supset SU(16)\times SU(3)_F\times U(1)_Z\supset 
SO(10)\times SU(3)_F\times U(1)_Z$,
an $SU(19)$ second-rank anti-symmetric tensor  representation 
${\bf 171}$ is decomposed into an $SU(16)$ second-rank anti-symmetric
tensor representation ${\bf 120}$, 
an $SU(3)_F$ defining representations ${\bf \overline{3}}$, and 
an $SU(16)\times SU(3)_F$ bi-fundamental representations 
$({\bf 16,3})$ given in Eq.~(\ref{branching-rule-171}).
Further, as is well-known, the $SO(10)$ spinor representation ${\bf 16}$
is decomposed into
$G_{\rm SM}\times U(1)_X=SU(3)_C\times SU(2)_L\times U(1)_Y\times U(1)_X$
representations:
\begin{align}
{\bf 16}=&
({\bf 3,2})(-1)(1)
\oplus({\bf \overline{3},1})(-2)(-3)
\oplus({\bf \overline{3},1})(4)(1)\nonumber\\
&\oplus({\bf 1,2})(3)(-3)
\oplus({\bf 1,1})(-6)(1)
\oplus({\bf 1,1})(0)(5).
\end{align}
That is, three generations of the SM Weyl fermions are embedded into
an 4D $SU(19)$ ${\bf 171}$ Weyl fermion.
In addition, an $SU(16)$ complex representation ${\bf 120}$ is
identified with an $SO(10)$ real representation ${\bf 120}$.
An 4D $SU(16)$ ${\bf 120}$ Weyl fermion is chiral,
while an 4D $SO(10)$ ${\bf 120}$ Weyl fermion is vectorlike.
When $SU(19)$ is broken to $SO(10)$,
$SU(3)_F\times U(1)_Z$ is broken to nothing. 
An 4D Weyl fermion in the $SO(10)\times SU(3)$
$({\bf 1},{\bf \overline{3}})$ is chiral, 
while three 4D $SO(10)$ ${\bf 1}$ Weyl fermion is vectorlike.
Thus, once $SU(19)$ is broken to $SO(10)$, an 4D $SU(19)$ ${\bf 191}$
Weyl fermion is decomposed into  three 4D $SO(16)$ ${\bf 16}$ Weyl
fermions and vectorlike fermions.

\begin{table}[t]
\begin{center}
\begin{tabular}{ccl}
 $SU(19)$
 &$\underset{\rm BCs}{\longrightarrow}$
 &$SU(16)\times SU(3)\times U(1)$\\
 &$\underset{\langle\Phi_{\bf 10830}\rangle\not=0}{\longrightarrow}$
 &$SO(10)\times SU(3)$\\
 &$\underset{\langle\Phi_{\bf 19}\rangle\mbox{\small 's}\not=0}{\longrightarrow}$
 &$SU(5)$\\
 &$\underset{\langle\Phi_{\bf 360}\rangle\not=0}{\longrightarrow}$
 &$SU(3)\times SU(2)\times U(1)=G_{\rm SM}$. 
\end{tabular}
\end{center}
 \caption{The table shows a symmetry breaking pattern of $SU(19)$ to
 $G_{\rm SM}$.
 BCs stands for a orbifold boundary condition.
 $\Phi_{\bf x}$  represents a scalar field in a representation ${\bf x}$
 of $SU(19)$.
 We assume that the appropriate component of each $\Phi_{\bf x}$
 develops its nonvanishing VEV $\langle\Phi_{\bf x}\rangle\not=0$.
\label{tab:SU19-Symmetry-Breaking-Pattern}}
\end{table}

Next, we consider a symmetry breaking pattern from $SU(19)$ to
$G_{\rm SM}$.  
One way of achieving it is to use orbifold symmetry breaking boundary
conditions (BCs) and several GUT breaking Higgses. One example is 
to choose orbifold BCs breaking $SU(19)$ to 
$SU(16)\times SU(3)_F\times U(1)$ and 
to introduce $SU(19)$ ${\bf 10830}$, ${\bf 360}$, 
${\bf 19}$ scalar fields, where we assume their proper components acquire
non-vanishing VEVs.
First, the following orbifold BC for the $SU(19)$ defining representation 
${\bf 19}$ breaks $SU(19)$ to $SU(16)\times SU(3)_F\times U(1)_Z$:
\begin{align}
P_{\bf 19}=\mbox{diag}\left(I_{16},-I_3\right),
\end{align}
where $P_{\bf 19}$ stands for a projection matrix defined in 
Eq.~(\ref{Eq:BC-SU(19)-boson-x}).
The non-vanishing VEV of the $SU(19)$ ${\bf 10830}$ scalar
field is responsible for breaking 
$SU(19)\supset SU(16)\times SU(3)_F\times U(1)_Z$ to
$SO(10)\times SU(3)_F$, where its branching rule of 
$SU(19)\supset SU(16)\times SU(3)_F\times U(1)_Z$ is given by
\begin{align}
{\bf 10830}=&
({\bf 5440,1})(12)
\oplus({\bf 1360,3})(-7)
\oplus({\bf 136,\overline{6}})(-26)
\nonumber\\
&\oplus({\bf 120,\overline{3}})(-26)
\oplus({\bf 16,8})(-45)
\oplus({\bf 1,6})(-64).
\label{branching-rule-10830}
\end{align}
An $SU(16)$ ${\bf 5440}$ contains singlet
under its $SO(10)$ special subgroup. 
Its nonvanishing VEV can break $SU(16)$ to its special subgroup
$SO(10)$ \cite{Yamatsu:2017sgu,Yamatsu:2017ssg}, where their $SO(10)$
decompositions are given in Ref.~\cite{Yamatsu:2015gut} by 
\begin{align}
{\bf 5440}=&
{\bf 4125}
\oplus{\bf \overline{1050}}
\oplus{\bf 210}
\oplus{\bf 54}
\oplus{\bf 1}.
\label{branching-rule-5440}
\end{align}
The VEV of an $SU(19)$ ${\bf 19}$ scalar breaks 
$(SU(19)\supset)SO(10)\times SU(3)_F\times  U(1)_Z$ to
$SU(5)\times SU(3)_F$ or
$(SU(19)\supset)SO(10)\times SU(3)_F\times  U(1)_Z$ to
$SO(10)\times SU(2)_F$, where its breaking rule is
given in Eq.~(\ref{branching-rule-19}).
If the three VEVs of the proper components of $SU(19)$ ${\bf 19}$
scalars can break $(SU(19)\supset)SO(10)\times SU(3)_F\times  U(1)_Z$ to
$SU(5)$.
The VEV of the $SU(19)$ ${\bf 360}$ scalar further
reduces $(SU(19)\supset SO(10)\times SU(3)_F\supset SO(10))\supset SU(5)$ 
to $G_{\rm SM}$, 
where its branching rule of 
$SU(19)\supset SU(16)\times SU(3)_F\times U(1)_Z$ is given by
\begin{align}
{\bf 360}=&
({\bf 255,1})(0)
\oplus({\bf 1,8})(0)
\oplus({\bf 1,1})(0)
\oplus({\bf 16,\overline{3}})(19)
\oplus({\bf \overline{16},3})(-19),
\label{branching-rule-360}
\end{align}
where the $SU(16)$ ${\bf 255}$
is decomposed into $SO(10)$ $({\bf 210}\oplus{\bf 45})$,
and the $SO(10)$ ${\bf 45}$ is decomposed into $SU(5)$
$({\bf 24}\oplus{\bf 10}\oplus{\bf \overline{10}}\oplus{\bf 1})$.
(For further information, see e.g., Ref.~\cite{Yamatsu:2015gut}.)

\section{$SU(19)$ special grand unification}
\label{Sec:Special-GUT}

As in Refs.~\cite{Yamatsu:2017sgu,Yamatsu:2017ssg}, 
an $SU(19)$ special GUT on 6D orbifold
spacetime $M^4\times T^2/\mathbb{Z}_2$ with 
the Randall-Sundrum (RS) type metric
\cite{Randall:1999ee,Hosotani:2017ghg,Hosotani:2017edv}
given by 
\begin{align}
ds^2=e^{-2\sigma(y)}(\eta_{\mu\nu}dx^{\mu}dx^{\nu}+dv^2)+dy^2,
\end{align}
where $y$ is the coordinate of RS warped space,
$v$ is the coordinate of $S^1$,
$\eta_{\mu\nu}=\mbox{diag}(-1,+1,+1,+1)$, 
$\sigma(y)=\sigma(-y)=\sigma(y+2\pi R_5)$,
$\sigma(y)=k|y|$ for $|y|\leq \pi R_5$, and
$v\sim v+2\pi R_6$.
There are four fixed points on $T^2/\mathbb{Z}_2$ 
at $(y_0,v_0)=(0,0)$, $(y_1,v_1)=(\pi R_5,0)$, 
$(y_2,v_2)=(0,\pi R_6)$, and $(y_3,v_3)=(\pi R_5,\pi R_6)$.
For each fixed point, the $\mathbb{Z}_2$ parity reflection 
is described by 
\begin{align}
P_j:\ (x_\mu,y_j+y,v_j+v)\ \to\ (x_\mu,y_j-y,v_j-v),
\end{align}
where $j=0,1,2,3$, and $P_3=P_1P_0P_2=P_2P_0P_1$.
5th and 6th dimensional translation 
$U_5: (x_\mu,y,v)\to(x_\mu,y+2\pi R_5,v)$ 
and $U_6: (x_\mu,y,v)\to(x_\mu,y,v+2\pi R_6)$  
satisfy $U_5=P_1P_0$ and $U_6=P_2P_0$, respectively.

We consider the matter content in the $SU(19)$ special GUT that consists
of a 6D $SU(19)$ bulk gauge boson $A_{M}$;
6D $SU(19)$ ${\bf 171}$ positive Weyl fermions with the orbifold BCs
$(\eta_{0},\eta_{1},\eta_{2},\eta_{3})=(+,+,-,-)$ and 
$(\eta_{0},\eta_{1},\eta_{2},\eta_{3})=(+,+,+,+)$ 
$\Psi_{{\bf 171}+}$ and $\Psi_{{\bf 171}+}'$, and 
6D negative Weyl fermions with $(-,+,-,+)$ and $(-,+,+,-)$
$\Psi_{{\bf 171}-}$ and $\Psi_{{\bf 171}-}'$, 
where $\eta_{j}$ $(j=0,1,2,3)$ stands for parity assignment for
each 6D fermion;
5D $SU(19)$, ${\bf 10830}$, ${\bf 360}$ and ${\bf 19}$ brane scalar
bosons at $y=0$ 
$\Phi_{\bf 10830}$, $\Phi_{\bf 360}$, $\Phi_{\bf 19}$,
$\Phi_{\bf 19}^{\prime(\alpha)}$ $(\alpha=1,2)$;
a 4D $SU(16)\times SU(3)_F\times U(1)_Z$ 
$({\bf \overline{120},1})(0)$ Weyl fermion,
four 4D $SU(16)\times SU(3)_F\times U(1)_Z$ 
$({\bf {120},1})(6)$ Weyl fermions,
sixty-four 4D $SU(16)\times SU(3)_F\times U(1)_Z$
$({\bf 1,{\bf \overline{3}}})(13)$ Weyl fermions,
four 4D $SU(16)\times SU(3)_F\times U(1)_Z$
$({\bf 1,{\bf \overline{3}}})(-32)$ Weyl fermions
at the fixed point $(y_0,v_0)=(0,0)$
$\psi_{\bf \overline{120}}$,
$\psi_{\bf {120}}^{(a)}$ $(a=1,2,3,4)$,
$\psi_{\bf \overline{3}}^{(b)}$ $(b=1,2,\cdots,64)$,
$\psi_{\bf \overline{3}}^{\prime(c)}$ $(c=1,2,3,4)$.
The matter content of the $SU(19)$ special GUT is
summarized in Table~\ref{tab:SU19-SU16-SO10-matter-content-6D}.
Note that the 5D brane scalars is responsible for achieving
appropriate symmetry and the 4D brane fermions are necessary to
realize 4D gauge anomaly cancellation. Only their conditions do not
uniquely determine the matter content, so one may choose another
matter content; e.g., one may introduce 6D bulk scalars instead of 5D
brane scalars. 
We will see the roles of the bulk and brane fields in the following.

\begin{table}[t]
\begin{center}
\begin{tabular}{cccccc}\hline
\rowcolor[gray]{0.8}
6D Bulk field&
 $A_M$&$\Psi_{{\bf 171}+}$&$\Psi_{{\bf 171}+}'$
 &$\Psi_{{\bf 171}-}$&$\Psi_{{\bf 171}-}'$
\\\hline
$SU(19)$ &${\bf 360}$&${\bf 171}$&${\bf 171}$&${\bf 171}$&${\bf 171}$\\
$SO(5,1)$&${\bf 6}$&${\bf 4}_+$&${\bf 4}_+$&${\bf 4}_-$&${\bf 4}_-$\\
Orbifold BC&
&$\left(
\begin{array}{cc}
-&-\\
+&+\\
\end{array}
 \right)$
 &$\left(
\begin{array}{cc}
+&+\\
+&+\\
\end{array}
 \right)$
 &$\left(
\begin{array}{cc}
-&+\\
-&+\\
\end{array}
\right)$
&$\left(
\begin{array}{cc}
+&-\\
-&+\\
\end{array}
\right)$\\
\hline
\end{tabular}\\[0.5em]
\begin{tabular}{ccccc}\hline
\rowcolor[gray]{0.8}
5D Brane field
&$\Phi_{\bf 10830}$&$\Phi_{\bf 360}$&$\Phi_{\bf 19}$&$\Phi_{\bf 19}^{\prime(\alpha)}$\\
\hline
$SU(19)$ &${\bf 10830}$&${\bf 360}$&${\bf 19}$&${\bf 19}$\\
$SO(4,1)$&{\bf 1}&{\bf 1}&{\bf 1}&{\bf 1}\\
Orbifold BC
&$\left(
\begin{array}{c}
-\\
-\\
\end{array}
\right)$
&$\left(
\begin{array}{c}
+\\
+\\
\end{array}
\right)$
&$\left(
\begin{array}{c}
+\\
+\\
\end{array}
\right)$
&$\left(
\begin{array}{c}
+\\
-\\
\end{array}
\right)$\\
Spacetime  &$y=0$&$y=0$&$y=0$&$y=0$\\
\hline
\end{tabular}\\[0.5em]
\begin{tabular}{ccccc}\hline
\rowcolor[gray]{0.8}
 4D field
 &$\psi_{{\bf \overline{120}}}$
 & $\psi_{{\bf {120}}}^{(a)}$
 &$\psi_{{\bf \overline{3}}}^{(b)}$
 &$\psi_{\overline{\bf 3}}^{\prime(c)}$
 \\
\hline
$SU(16)\times SU(3)_F$
 &$({\bf \overline{120},1})$
 &$({\bf {120},1})$
 &$({\bf 1,\overline{3}})$
 &$({\bf 1,\overline{3}})$\\	     
$U(1)_Z$  &$0$&$6$&$13$&$-32$\\
$SL(2,\mathbb{C})$&$(1/2,0)$&$(1/2,0)$&$(1/2,0)$&$(1/2,0)$\\
Spacetime $(y,v)$ &$(0,0)$&$(0,0)$&$(0,0)$&$(0,0)$\\
\hline
\end{tabular}
\end{center}
\caption{The table shows the matter content in
the $SU(19)$ special GUT on $M^4\times T^2/\mathbb{Z}_2$.
The representations of $SU(19)$ and 6D, 5D, 4D Lorentz group, 
the orbifold BCs of 6D bulk fields and 5D brane fields, and 
the spacetime location of 5D and 4D fields
 are shown.
Orbifold BCs stand for parity assignment \usebox{\BCmatrixA}
 for 6D  fields and \usebox{\BCmatrixB} for 5D fields.
 The orbifold BCs of the 6D $SU(19)$ gauge field $A_M$ are given in
 Eqs.~(\ref{Eq:gauge-field-BCs}) and (\ref{Eq:SU19-BCs-19}).
 ($\alpha=1,2$;
 $a=1,2,3,4$;
 $b=1,2,\cdots,64$;
 $c=1,2,3,4$.)
\label{tab:SU19-SU16-SO10-matter-content-6D}}
\end{table}

First, a 6D $SU(19)$ bulk gauge boson $A_{M}$ is decomposed into 
a 4D gauge field $A_\mu$ and 5th and 6th dimensional gauge fields $A_y$
and $A_v$. The orbifold BCs of the 6D $SU(19)$ gauge field are given by
\begin{align}
\left(
\begin{array}{c}
A_\mu\\
A_y\\
A_v\\
\end{array}
\right)(x,y_j-y,v_j-v)
=P_{j{\bf 19}}
\left(
\begin{array}{c}
A_\mu\\
-A_y\\
-A_v\\
\end{array}
\right)(x,y_j+y,v_j+v)
P_{j{\bf 19}}^{-1},
\label{Eq:gauge-field-BCs}
\end{align}
where $P_{j{\bf 19}}$ is a projection matrix satisfying 
$(P_{j{\bf 19}})^2=I_{19}$. 
We consider the orbifold BCs $P_2$ and $P_3$ preserving $SU(19)$
symmetry, while the orbifold BCs $P_0$ and $P_1$ reduce $SU(19)$ to its
regular subgroup $SU(16)\times SU(3)_F\times U(1)_Z$.
We take $P_{j{\bf 19}}$ as 
\begin{align}
P_{j{\bf 19}}=
\left\{
\begin{array}{ll}
I_{19}&\mbox{for}\ j=2,3\\
\mbox{diag}\left(I_{16},-I_{3}\right)  &\mbox{for}\ j=0,1
\end{array}
\right..
\label{Eq:SU19-BCs-19}
\end{align}
In this case, the 4D $SU(19)$ ${\bf 360}$ gauge field $A_\mu$ have
Neumann BCs at the fixed points $(y_2,v_2)$ and 
$(y_3,v_3)$, while the 5th and 6th dimensional gauge
fields $A_y$ and $A_v$ have Dirichlet BCs because of the negative sign
in Eq.~(\ref{Eq:gauge-field-BCs}).
On the other hand, since $SU(19)$ symmetry is broken to 
$SU(16)\times SU(3)_F\times U(1)_Z$ at the fixed points $(y_0,v_0)$ and
$(y_1,v_1)$, 
by using the branching rules of the $SU(19)$ adjoint representation 
${\bf 360}$ given in Eq.~(\ref{branching-rule-360})
as well as  Eqs.~(\ref{Eq:gauge-field-BCs}) and (\ref{Eq:SU19-BCs-19}),
the $SU(16)\times SU(3)_F \times U(1)_Z$
$\left(({\bf 255,1})(0)\oplus({\bf 1,8})(0)\oplus({\bf 1,1})(0)\right)$
and $\left(({\bf 16,\overline{3}})(19)
\oplus({\bf \overline{16},3})(-19)\right)$ 
components of the 4D gauge field $A_\mu$ 
have Neumann and Dirichlet BCs at
the fixed points $(y_0,v_0)$ and $(y_1,v_1)$, respectively;
the $SU(16)\times SU(3)_F\times U(1)_Z$ 
$\left(({\bf 255,1})(0)\oplus({\bf 1,8})(0)\oplus({\bf 1,1})(0)\right)$
and $\left(({\bf 16,\overline{3}})(19)
\oplus({\bf \overline{16},3})(-19)\right)$ 
components of the 5th and 6th dimensional gauge 
fields $A_y$ and $A_v$ have Dirichlet and Neumann BCs, respectively.
Thus, since the $SU(16)\times SU(3)_F\times U(1)_Z$ 
$\left(({\bf 255,1})(0)\oplus({\bf 1,8})(0)\oplus({\bf 1,1})(0)\right)$
components of the 4D gauge field $A_\mu$ have four Neumann BCs at the
four fixed points $(y_j,v_j) (j=0,1,2,3)$, they have zero modes
corresponding to 4D $SU(16)$, $SU(3)_F$ and $U(1)_Z$ gauge fields;
since the other components of $A_\mu$ and any component of $A_y$ and
$A_v$ have four Dirichlet BCs or two Neumann and two Dirichlet BCs at the
four fixed points, they do not have zero modes.
The orbifold BCs reduce $SU(19)$ to $SU(16)\times SU(3)_F\times U(1)_Z$. 

To achieve the SM gauge symmetry $G_{\rm SM}$ at low energies, 
we consider the symmetry breaking sector via spontaneous symmetry
breaking. We introduce 5D $SU(19)$ ${\bf 10830}$, ${\bf 360}$ and
${\bf 19}$ brane scalar fields,
$\Psi_{\bf 10830}$, $\Phi_{\bf 360}$, $\Phi_{\bf 19}$, and
$\Phi_{\bf 19}^{\prime(\alpha)}$ $(\alpha=1,2)$ on the 5D brane ($y=0$).
Their orbifold BCs are given by 
\begin{align}
\Phi_{\bf x}^{(\prime)}(x,v_\ell-v)=&
 \eta_{\ell{\bf x}}^{(\prime)}P_{\ell{\bf x}}\Phi_{\bf x}^{(\prime)}(x,v_\ell+v),
\label{Eq:BC-SU(19)-boson-x}
\end{align}
where $\ell=0,2$, ${\bf x}$ stands for
${\bf 10830}$, ${\bf 360}$ and ${\bf 19}$,
$\eta_{\ell{\bf x}}$ is a positive or negative sign, and
$P_{\ell{\bf x}}$ is a projection matrix. 
We take $\eta_{\ell{\bf 10830}}=-\eta_{\ell{\bf 360}}
=-\eta_{\ell{\bf 19}}=\eta_{0{\bf 19}}'=-\eta_{2{\bf 19}}'=-1$.
The branching rules of $SU(19)\supset SU(16)\times SU(3)_F\times U(1)_Z$
for ${\bf 10830}$, ${\bf 360}$ and ${\bf 19}$ are given in
Eqs.~(\ref{branching-rule-10830}), (\ref{branching-rule-360}), and
(\ref{branching-rule-19}), respectively.
For $\Phi_{\bf 10830}$, the $SU(16)\times SU(3)_F\times U(1)_Z$
$({\bf 5440,1})(12)\oplus({\bf 120,\overline{3}})(-26)
\oplus({\bf 1,6})(-64)$ components have zero modes;
for $\Phi_{\bf 360}$, the $SU(16)\times SU(3)_F\times U(1)_Z$
$({\bf 255,1})(0)\oplus({\bf 1,8})(0)\oplus({\bf 1,1})(0)$
components have zero modes;
for $\Phi_{\bf 19}$, the $SU(16)\times SU(3)_F\times U(1)_Z$ 
$({\bf 16,1})(3)$ components have zero modes,
and for $\Phi_{\bf 19}^{\prime(\alpha)}$, the
$SU(16)\times SU(3)_F\times U(1)_Z$ 
$({\bf 1,3})(-16)$ components have zero modes.
We assume that a scalar field $\Phi_{\bf 10830}$ is responsible for
breaking  $(SU(19)\supset)SU(16)\times SU(3)_F\times U(1)_Z$ to
$SO(10)\times SU(3)_F$;
two scalar fields $\Phi_{\bf 19}^{\prime(\alpha)}$ are responsible for
breaking  $(SU(19)\supset SU(16)\times SU(3)_F\times U(1)_Z)
\supset SO(10)\times SU(3)_F$ to $SO(10)$;
the nonvanishing VEV of the scalar field 
$\Phi_{\bf 19}$ breaks $(SU(19)\supset)SO(10)$ to $SU(5)$;
the nonvanishing VEV of $\Phi_{\bf 360}$ breaks $(SU(19)\supset)SU(5)$
to $G_{\rm SM}$.

The SM Weyl fermions are identified with zero modes of a 6D $SU(19)$ 
${\bf 171}$ Weyl bulk fermion. The orbifold BCs of 6D $SU(19)$ 
${\bf 171}$ positive or negative Weyl bulk fermions can be written by 
\begin{align}
\Psi_{{\bf 171}\pm}^{(\prime)}(x,y_j-y,v_j-v)&=
\eta_{j{\bf 171}\pm}^{(\prime)}\overline{\gamma}
P_{j{\bf 171}}
\Psi_{{\bf 171}\pm}^{(\prime)}
(x,y_j+y,v_j+v),
\label{Eq:BC-SU(19)-fermion-171}
\end{align}
where the subscript of $\Psi$ $\pm$ stands for 6D chirality,
$\eta_{j{\bf 171}\pm}^{(\prime)}$ is a positive or negative sign,
$\prod_{j=0}^{3}\eta_{j{\bf 171}\pm}^{(\prime)}=1$,
6D gamma matrices $\gamma^a$ $(a=1,2,\cdots,7)$ 
satisfy $\{\gamma^a,\gamma^b\}=2\eta^{ab}$
($\eta^{ab}=\mbox{diag}(-I_1,I_5)$),
$\overline{\gamma}
:=-i\gamma^5\gamma^6
=\gamma_{\rm 6D}^7\gamma_{\rm 4D}^5$,
$\gamma_{\rm 4D}^5=I_2\otimes \sigma^3\otimes I_2$,
$\gamma_{\rm 6D}^7=I_4 \otimes \sigma^3$, 
and $P_{j{\bf 171}\pm}^{(\prime)}$ is a projection matrix.
(The same notation is used in
Refs.~\cite{Hosotani:2017ghg,Hosotani:2017edv}.)
In our notation, a 6D Dirac fermion $\Psi_{\rm D}^{\rm 6D}$
and 6D positive and negative Weyl fermions
$\Psi_{\pm}^{\rm 6D}:=P_{\pm}^{\rm 6D}\Psi_{\rm D}^{\rm 6D}$
($P_{\pm}^{\rm 6D}:=(1\pm\gamma_{\rm 6D}^7)/2$)
can be expressed by using 4D left- and right-handed Weyl fermions
$\psi_{L/R\pm}^{\rm 4D}(=P_{L/R}^{\rm 4D}\psi_{\rm D}^{\rm 4D})$
($P_{L/R}^{\rm 4D}:=(1\pm\gamma_{\rm 4D}^5)/2$),
where subscripts $L/R$ and $\pm$ stand for
4D and 6D chiralities, respectively:
\begin{align}
 &\Psi_{\rm D}^{\rm 6D}
 :=
 \left(
 \begin{array}{c}
  \psi_{R+}^{\rm 4D}\\
  \psi_{L+}^{\rm 4D}\\
  \psi_{R-}^{\rm 4D}\\
  \psi_{L-}^{\rm 4D}\\
 \end{array}
 \right),\ \
 \Psi_{+}^{\rm 6D}
 =P_{+}^{\rm 6D}\Psi_{\rm D}^{\rm 6D}
 =\left(
 \begin{array}{c}
  \psi_{R+}^{\rm 4D}\\
  \psi_{L+}^{\rm 4D}\\
  0\\
  0\\
 \end{array}
 \right),\ \
 \Psi_{-}^{\rm 6D}
 =P_{-}^{\rm 6D}\Psi_{\rm D}^{\rm 6D}
 = \left(
 \begin{array}{c}
  0\\
  0\\
  \psi_{R-}^{\rm 4D}\\
  \psi_{L-}^{\rm 4D}\\
 \end{array}
 \right).
\end{align}
Note that to see the relation between $P_{j{\bf 171}}$ and
$P_{j{\bf 19}}$, we express the orbifold BCs of
$\Psi_{{\bf 171}\pm}^{(\prime)}$ by using a $19\times 19$
matrix form as the same as one of the gauge field in
Eq.~(\ref{Eq:gauge-field-BCs}).
We can write the $(\alpha\beta)$ component of 
$\Psi_{{\bf 171}\pm}^{(\prime)}$ as 
$[\Psi_{{\bf 171}\pm}^{(\prime)}]_{\alpha\beta}
=\sum_{a<b}\psi_{{\bf 171}\pm}^{(\prime)ab}
[M_{{\bf 171}ab}]_{\alpha\beta}$
$(a,b,\alpha,\beta=1,2,\cdots,19)$,
where $\psi_{{\bf 171}\pm}^{(\prime)ab}$s are the component fields
of $\Psi_{{\bf 171}\pm}^{(\prime)}$ expanded by
$M_{{\bf 171}ab}$s, and $[M_{{\bf 171}ab}]_{\alpha\beta}=(1/\sqrt{2})
\left(\delta_{a\alpha}\delta_{b\beta}-\delta_{a\beta}\delta_{b\alpha}\right)$.
In this notation, the orbifold BCs of the 6D $SU(19)$ anti-symmetric
tensor fermion $\Psi_{{\bf 171}\pm}^{(\prime)}$ can be expressed by
using the projection matrix $P_{j{\bf 19}}$
given in Eq.~(\ref{Eq:SU19-BCs-19})
instead of $P_{j{\bf 171}}$:
\begin{align}
[\Psi_{{\bf 171}\pm}^{(\prime)}(x,y_j-y,v_j-v)]_{\alpha\beta}&=
\eta_{j{\bf 171}\pm}^{(\prime)}\overline{\gamma}
 [P_{j{\bf 19}}]_{\alpha\kappa}
 [\Psi_{{\bf 171}\pm}^{(\prime)}(x,y_j+y,v_j+v)]_{\kappa\lambda}
 [P_{j{\bf 19}}]_{\lambda\beta},
\label{Eq:BC-SU(19)-fermion-171-2}
\end{align}
where $\alpha,\beta,\kappa,\lambda=1,2,\cdots,19$;
$[P_{j{\bf 19}}]_{\alpha\beta}$ denotes the $(\alpha\beta)$ element of
the projection matrix $P_{j{\bf 19}}$ given in
Eq.~(\ref{Eq:SU19-BCs-19}).
(The projection matrix of any $SU(19)$ tensor product representation can
be expressed by the projection matrix of the $SU(19)$ defining
representation $P_{j{\bf 19}}$.)

Here, we check zero modes of, e.g., a 6D $SU(19)$ ${\bf 171}$ positive
Weyl fermion with orbifold BCs
$(\eta_{0},\eta_{1},\eta_{2},\eta_{3})=(+,+,-,-)$ $\Psi_{{\bf 171}+}$.
At fixed points $(y_2,v_2)$ and $(y_3,v_3)$, 
the 4D $SU(19)$ ${\bf 171}$ left-handed Weyl fermion components have
Neumann BCs, while the 4D $SU(19)$ ${\bf 171}$ right-handed Weyl fermion
components have Dirichlet BCs.
At fixed points $(y_0,v_0)$ and $(y_1,v_1)$, 
the 4D $SU(16)\times SU(3)_F\times U(1)_Z$
$({\bf 16},{\bf 3})(-13)$ and
$({\bf 120},{\bf 1})(6)\oplus({\bf 1},{\bf \overline{3}})(-32)$ 
left-handed Weyl fermion components have Neumann and Dirichlet BCs,
respectively, 
while the 4D $SU(16)\times SU(3)_F\times U(1)_Z$
$({\bf 16},{\bf 3})(-13)$ and
$({\bf 120},{\bf 1})(6)\oplus({\bf 1},{\bf \overline{3}})(-32)$ 
right-handed Weyl fermion components have Dirichlet and Neumann BCs,
respectively. 
In this case, only the 4D $SU(16)\times SU(3)_F\times U(1)_Z$
$({\bf 16},{\bf 3})(-13)$ 
left-handed Weyl fermion has zero modes.
Also, for $\Psi_{{\bf 171}+}^{\prime}$,
the 4D $SU(16)\times SU(3)_F\times U(1)_Z$
$({\bf 120},{\bf 1})(6)\oplus({\bf 1},{\bf \overline{3}})(-32)$
right-handed Weyl fermion has zero modes.
They are vectorlike once we take into account
the symmetry breaking effects of $SU(16)\times SU(3)_F\times U(1)_Z$
to $SO(10)$.
Also, for $\Psi_{{\bf 171}-}$ and $\Psi_{{\bf 171}-}^{\prime}$,
there is no zero mode.
The parity assignments of $\Psi_{{\bf 171}\pm}^{(\prime)}$ are
summarized in Table~\ref{Tab:parity-assignment-fermion-171}.

\begin{table}[t]
\begin{center}
\begin{tabular}{ccccc}\hline
 \rowcolor[gray]{0.8}
&\multicolumn{2}{c}{$\Psi_{{\bf 171}+}$}
&\multicolumn{2}{c}{$\Psi_{{\bf 171}+}'$}
\\
\rowcolor[gray]{0.8}  
$SU(16)\times SU(3)\times U(1)$&Left&Right&Left&Right\\\hline
$({\bf 16,3})(-13)$&
$\left(
\begin{array}{cc}
+&+\\
+&+\\
\end{array}
\right)$&
$\left(
\begin{array}{cc}
-&-\\
-&-\\
\end{array}
\right)$
&$\left(
\begin{array}{cc}
-&-\\
+&+\\
\end{array}
\right)$&
$\left(
\begin{array}{cc}
+&+\\
-&-\\
\end{array}
\right)$\\
$({\bf 120,1})(6)$&
$\left(
\begin{array}{cc}
+&+\\
-&-\\
\end{array}
\right)$&
$\left(
\begin{array}{cc}
-&-\\
+&+\\
\end{array}
\right)$
&$\left(
\begin{array}{cc}
-&-\\
-&-\\
\end{array}
\right)$&
$\left(
\begin{array}{cc}
+&+\\
+&+\\
\end{array}
\right)$\\
$({\bf 1,\overline{3}})(-32)$&
$\left(
\begin{array}{cc}
+&+\\
-&-\\
\end{array}
\right)$&
$\left(
\begin{array}{cc}
-&-\\
+&+\\
\end{array}
\right)$
&$\left(
\begin{array}{cc}
-&-\\
-&-\\
\end{array}
\right)$&
$\left(
\begin{array}{cc}
+&+\\
+&+\\
\end{array}
\right)$\\
\hline
 \end{tabular}\\[0.5em]
 \begin{tabular}{ccccc}\hline
  \rowcolor[gray]{0.8}
&\multicolumn{2}{c}{$\Psi_{{\bf 171}-}$}
&\multicolumn{2}{c}{$\Psi_{{\bf 171}-}'$}
\\
\rowcolor[gray]{0.8}  
$SU(16)\times SU(3)\times U(1)$&Left&Right&Left&Right\\\hline
$({\bf 16,3})(-13)$&
$\left(
\begin{array}{cc}
-&+\\
+&-\\
\end{array}
\right)$&
$\left(
\begin{array}{cc}
+&-\\
-&+\\ 
\end{array}
\right)$
&$\left(
\begin{array}{cc}
+&-\\
+&-\\
\end{array}
\right)$&
$\left(
\begin{array}{cc}
-&+\\
-&+\\
\end{array}
\right)$\\
$({\bf 120,1})(6)$&
$\left(
\begin{array}{cc}
-&+\\
-&+\\
\end{array}
\right)$&
$\left(
\begin{array}{cc}
+&-\\
+&-\\
\end{array}
\right)$
&$\left(
\begin{array}{cc}
+&-\\
-&+\\
\end{array}
\right)$&
$\left(
\begin{array}{cc}
-&+\\
+&-\\
\end{array}
\right)$\\
$({\bf 1,\overline{3}})(-32)$&
$\left(
\begin{array}{cc}
-&+\\
-&+\\
\end{array}
\right)$&
$\left(
\begin{array}{cc}
+&-\\
+&-\\
\end{array}
\right)$
&$\left(
\begin{array}{cc}
+&-\\
-&+\\
\end{array}
\right)$&
$\left(
\begin{array}{cc}
-&+\\
+&-\\
\end{array}
\right)$\\
\hline
 \end{tabular}
 \caption{The table shows parity assignments \usebox{\BCmatrix}
 of the 4D $SU(16)\times SU(3)\times U(1)$ left- and right-handed Weyl
 fermion components 
 of the 6D $SU(19)$ ${\bf 171}$ positive and negative Weyl fermions
 $\Psi_{{\bf 171}+}$,
 $\Psi_{{\bf 171}+}^{\prime}$,
 $\Psi_{{\bf 171}-}$, and 
 $\Psi_{{\bf 171}-}^{\prime}$
 given in Eq.~(\ref{Eq:BC-SU(19)-fermion-171}).
 }
\label{Tab:parity-assignment-fermion-171}
\end{center}
\end{table}

Here, we check the contribution to 6D bulk and 4D brane anomalies from
the above 6D Weyl fermion sets. 
The fermion set does not contribute to 6D $SU(19)$ gauge anomaly because
of the same number of 6D $SU(19)$ ${\bf 171}$ positive and negative Weyl
fermions.
We need to check 4D gauge anomaly cancellation at four fixed points 
$(y_j,v_j) (j=0,1,2,3)$ by using 4D anomaly coefficients listed
in Ref.~\cite{Yamatsu:2015gut}.
From Table~\ref{Tab:parity-assignment-fermion-171},
at three fixed points $(y_j,v_j) (j=1,2,3)$,
there are two 4D left- and right-handed Weyl fermions in
$({\bf 16,3})(-13)$, $({\bf 120,1})(6)$, and
$({\bf 1,\overline{3}})(-32)$ of $SU(16)\times SU(3)_F\times U(1)$
from the 6D $SU(19)$ ${\bf 171}$ positive and negative Weyl
fermions $\Psi_{{\bf 171}+}$, $\Psi_{{\bf 171}+}'$,
$\Psi_{{\bf 171}-}$, and $\Psi_{{\bf 171}-}'$.
The vectorlike matter sets do not produce any 4D gauge anomalies.
At the other fixed point $(y_0,v_0)$, there can be 4D pure $SU(16)$,
pure $SU(3)_F$ pure $U(1)_Z$, mixed $SU(16)-SU(16)-U(1)_Z$, mixed
$SU(3)_F-SU(3)_F-U(1)_Z$, and mixed $\mbox{grav.}-\mbox{grav.}-U(1)_Z$
anomalies. In fact, the 6D $SU(19)$ 
${\bf 171}$ positive and negative Weyl fermions generate 
4D pure $SU(16)$, pure $SU(3)_F$, pure $U(1)_Z$,
mixed $SU(16)-SU(16)-U(1)_Z$, mixed $SU(3)_F-SU(3)_F-U(1)_Z$,
and mixed $\mbox{grav.}-\mbox{grav.}-U(1)_Z$
anomalies. We focus on how to cancel the 4D anomalies at the fixed point
$(y_0,v_0)$ below.

To achieve 4D gauge anomaly cancellation at the fixed point $(y_0,v_0)$,
we need to introduce 4D Weyl fermions in appropriate
representations of $SU(16)\times SU(3)_F\times U(1)$.
First, we consider the pure $SU(16)$ gauge anomaly cancellation.
The 4D $SU(16)$ gauge anomaly of twelve 4D $SU(16)$ ${\bf 16}$
left-handed Weyl fermions
and four 4D $SU(16)$ ${\bf 120}$ right-handed Weyl fermions
is canceled out by the anomaly of
three 4D $SU(16)$ ${\bf 120}$ left-handed Weyl fermions.
From Table~\ref{tab:SU19-SU16-SO10-matter-content-6D},
there are one 4D $SU(16)$ ${\bf \overline{120}}$ left-handed Weyl fermion 
$\psi_{{\bf \overline{120}}}$
and four 4D $SU(16)$ ${\bf 120}$ left-handed Weyl fermions
$\psi_{{\bf {120}}}^{(a)}$ $(a=1,2,3,4)$ in the model.
The $SU(16)$ gauge anomaly of $\psi_{{\bf \overline{120}}}$ is
canceled out by one of e.g., $\psi_{{\bf {120}}}^{(1)}$.
The other three 4D $SU(16)$ ${\bf 120}$ left-handed Weyl fermions
$\psi_{{\bf {120}}}^{(a)}$ $(a=2,3,4)$ contribute the $SU(16)$ gauge
anomaly on the fixed point $(y_0,v_0)$.
Thus, the 4D brane fermions $\psi_{{\bf \overline{120}}}$
and $\psi_{{\bf {120}}}^{(a)}$ $(a=1,2,3,4)$ cancel 
the 4D $SU(16)$ gauge anomaly from the bulk fermions.
Second, for the pure $SU(3)_F$ gauge anomaly, 
the 4D $SU(3)$ gauge anomaly of sixty-four 4D $SU(3)$ ${\bf 3}$
left-handed Weyl fermions
and four 4D $SU(3)$ ${\bf \overline{3}}$ right-handed Weyl fermions
is canceled out by the anomaly of sixty-eight 4D $SU(3)$
${\bf \overline{3}}$ left-handed Weyl fermions.
Third, for 4D pure $U(1)_Z$, mixed $SU(16)-SU(16)-U(1)_Z$,
mixed $SU(3)_F-SU(3)_F-U(1)_Z$, and mixed
$\mbox{grav.}-\mbox{grav.}-U(1)_Z$ anomalies are canceled out 
if the matter content is vectorlike from the view of the $U(1)_Z$
gauge theory.
The matter content shown in
Table~\ref{tab:SU19-SU16-SO10-matter-content-6D}
satisfies all the above requirements, so
any 6D and 4D gauge anomalies at the fixed points are canceled out.

\section{Summary and discussion}
\label{Sec:Summary-discussion}

In this paper, we have pointed out that in special GUT framework, family
unification may be achieved by using GUT groups and their
``regular-type'' and ``product-type'' subgroups, such as 
$SU(19)\supset SU(16)\times SU(3)\times U(1)$ and 
$SU(48)\supset SU(16)\times SU(3)$, respectively.

For a ``regular-type'' subgroup
$SU(19)\supset SU(16)\times SU(3)\times U(1)$, 
we have constructed an $SU(19)$ special GUT by using a
special breaking $SU(16)$ to $SO(10)$. In this framework, the zero modes
of a 6D $SU(19)$ ${\bf 171}$ Weyl fermion can be identified with
three generations of quarks and leptons;
the 6D $SU(19)$ gauge anomaly on the bulk and the 4D $SU(19)$ or
$SU(16)\times SU(3)\times U(1)$ gauge anomalies at each fixed point can
be canceled out; as in the $SU(16)$ special GUT
\cite{Yamatsu:2017sgu},  exotic 
chiral 
fermions do not exist due to a special feature of 
the $SU(16)$ complex representation ${\bf \overline{120}}$ once we take
into account the symmetry breaking of $SU(19)$ to $SO(10)$.

To cancel 4D pure $SU(16)$, pure $SU(3)_F$, pure $U(1)_Z$, and mixed
anomalies on a fixed point, we introduced a lot of 4D Weyl fermions.
For the mixed anomalies, one may rely on Green-Schwarz (GS) anomaly
cancellation mechanism \cite{Green1984} for 4D version
\cite{Binetruy:1996uv,Kojima:2017qbt}.
It may be achieved by introducing a pseudo-scalar field that
transforms non-linearly under the anomalous $U(1)_Z$ symmetry.
In this case, the number of 4D Weyl fermions can be reduced, drastically.

We comment on the SM fermion masses in the $SU(19)$ special GUT. Since
three generations of the SM fermions are unified into a 6D $SU(19)$
${\bf 171}$ Weyl fermion, the masses of all quarks and leptons are
degenerate without 
$SU(19)(\supset SU(16)\times SU(3)_F\supset SO(10)\times SU(3)_F)$
breaking effects. 
We assumed that since the nonvanishing VEVs of 5D brane scalars
$\Psi_{\bf 19}^{\prime(\alpha)}$ break the $SU(3)_F$ symmetry,
there is no reason to expect the unified masses of first, second and
third generations of up-type and down-type quarks, charged leptons, and
neutrinos, respectively; in addition, since the nonvanishing VEVs of 5D
brane scalars $\Psi_{\bf 10830}$, $\Psi_{\bf 19}$, and $\Psi_{\bf 360}$
break $SU(16)(\supset SO(10))$ to $G_{\rm SM}$,
there is no reason to expect the degenerate mass of quarks and leptons
for each generation.
As discussed in e.g.,
Refs.~\cite{Hosotani:2017ghg,Hosotani:2017edv}, on the UV brane $y=0$,
we can introduce $SU(19)$-invariant brane interaction terms among the 6D
bulk fermions and the 5D brane scalars because $SU(19)$ tensor products
e.g., ${\bf 171}\otimes{\bf 171}\otimes{\bf \overline{10830}}$,
${\bf 171}\otimes{\bf \overline{171}}\otimes{\bf 360}$,
${\bf 171}\otimes{\bf \overline{171}}\otimes
{\bf 19}\otimes{\bf \overline{19}}$,
etc. contain singlet.
The $SU(16)\times SU(3)_F\times U(1)_Z$ $({\bf 16,3})(-13)$
and $({\bf 120,1})(6)$ components of the $SU(19)$ ${\bf 171}$ bulk
fermions can be mixed via the VEVs of the 5D brane scalars once their
corresponding brane interaction terms or effective brane mass terms are 
generated, where $SO(10)(\subset SU(19))$ ${\bf 120}$ contains
${({\bf 3,2})(-1)}$, ${({\bf \overline{3},1})(4)}$, 
${({\bf \overline{3},1})(-2)}$, ${({\bf 1,2})(3)}$, 
and ${({\bf 1,1})(0)}$ of $G_{\rm SM}$.
We expect that the effective mass terms divide the degenerate mass for
each generation into up-type quark, down-type quark, charged lepton, and 
neutrino masses, where some VEVs or coupling constants must be
hierarchical to realize mass hierarchies for up-type and down-type
quarks and charged leptons. To realize tiny neutrino masses, it seems to
be better to introduce 5D symplectic Majorana fermions
\cite{Mirabelli:1997aj} on the UV brane $y=0$. $SU(19)$ brane 
interaction terms among the 6D bulk fermions, the 5D brane scalars, and
the 5D symplectic Majorana fermions lead to tiny neutrino masses via a
seesaw mechanism discussed in
Refs.~\cite{Hosotani:2017ghg,Hasegawa:2018jze}. 
The above brane interaction terms are essential to realize not only
quark and lepton masses but also their mixing matrices, i.e., 
the Cabibbo-Kobayashi-Maskawa (CKM)
\cite{Cabibbo:1963yz,Kobayashi:1973fv} and
Maki-Nakagawa-Sakata (MNS) \cite{Maki:1962mu} matrices.
Since we can introduce a lot of 5D brane interaction terms in the
$SU(19)$ special GUT, the model seems to realize the SM fermion
masses, but seems to give us no prediction about quark and lepton
masses and mixings. We will leave the detail analysis in future studies.

We have discussed how to embed three chiral generations of quarks
and leptons in triplet (a finite-dimensional representation) of a
non-Abelian compact group $SU(3)_F$.
Another direction for unifying generations may be considered by using
non-Abelian noncompact groups (e.g., $SU(1,1)$) and their
infinite-dimensional representation
\cite{Inoue:1994qz,Inoue:2000ia,Inoue:2003qi,Yamatsu:2007,Yamatsu:2008,Yamatsu:2009,Yamatsu:2012,Yamatsu:2013}.

\section*{Acknowledgments}

The author would like to thank Kenji Nishiwaki
for critical reading of the manuscript and many valuable comments.

\bibliographystyle{utphys} 
\bibliography{../../arxiv/reference}

\providecommand{\href}[2]{#2}\begingroup\raggedright\begin{thebibliography}{10}

\bibitem{Wilczek:1978xi}
F.~Wilczek and A.~Zee, ``{Horizontal Interaction and Weak Mixing Angles},''
\href{http://dx.doi.org/10.1103/PhysRevLett.42.421}{{\em Phys. Rev. Lett.}
  {\bfseries 42} (1979) 421}.

\bibitem{Froggatt:1978nt}
C.~D. Froggatt and H.~B. Nielsen, ``{Hierarchy of Quark Masses, Cabibbo Angles
  and CP Violation},''
\href{http://dx.doi.org/10.1016/0550-3213(79)90316-X}{{\em Nucl. Phys.}
  {\bfseries B147} (1979) 277}.

\bibitem{Yanagida:1979as}
T.~Yanagida, ``{Horizontal Gauge Symmetry and Masses of Neutrinos},''. In
  Proceedings of the Workshop on the Baryon Number of the Universe and Unified
  Theories, Tsukuba, Japan, p95 (1979).

\bibitem{Maehara:1979kf}
T.~Maehara and T.~Yanagida, ``{Gauge Symmetry of Horizontal Flavor},''
\href{http://dx.doi.org/10.1143/PTP.61.1434}{{\em Prog. Theor. Phys.}
  {\bfseries 61} (1979) 1434}.

\bibitem{Inoue:1994qz}
K.~Inoue, ``{Generations of Quarks and Leptons from Noncompact Horizontal
  Symmetry},'' \href{http://dx.doi.org/10.1143/PTP.93.403}{{\em Prog. Theor.
  Phys.} {\bfseries 93} (1995) 403--416},
\href{http://arxiv.org/abs/hep-ph/9410220}{{\ttfamily arXiv:hep-ph/9410220}}.

\bibitem{King:2001uz}
S.~F. King and G.~G. Ross, ``{Fermion Masses and Mixing Angles from $SU(3)$
  Family Symmetry},''
  \href{http://dx.doi.org/10.1016/S0370-2693(01)01139-X}{{\em Phys. Lett.}
  {\bfseries B520} (2001) 243--253},
\href{http://arxiv.org/abs/hep-ph/0108112}{{\ttfamily arXiv:hep-ph/0108112}}.

\bibitem{Maekawa2004}
N.~Maekawa and T.~Yamashita, ``{Horizontal Symmetry in Higgs Sector of GUT with
  $U(1)_A$ Symmetry},''
  \href{http://dx.doi.org/10.1088/1126-6708/2004/07/009}{{\em JHEP} {\bfseries
  07} (2004) 009},
\href{http://arxiv.org/abs/hep-ph/0404020}{{\ttfamily arXiv:hep-ph/0404020}}.

\bibitem{Yoshioka:1999ds}
K.~Yoshioka, ``{On Fermion Mass Hierarchy with Extra Dimensions},''
  \href{http://dx.doi.org/10.1142/S0217732300000062,
  10.1016/S0217-7323(00)00006-2}{{\em Mod. Phys. Lett.} {\bfseries A15} (2000)
  29--40},
\href{http://arxiv.org/abs/hep-ph/9904433}{{\ttfamily arXiv:hep-ph/9904433
  [hep-ph]}}.

\bibitem{Fujimoto:2012wv}
Y.~Fujimoto, T.~Nagasawa, K.~Nishiwaki, and M.~Sakamoto, ``{Quark Mass
  Hierarchy and Mixing via Geometry of Extra Dimension with Point
  Interactions},'' \href{http://dx.doi.org/10.1093/ptep/pts097}{{\em PTEP}
  {\bfseries 2013} (2013) 023B07},
\href{http://arxiv.org/abs/1209.5150}{{\ttfamily arXiv:1209.5150 [hep-ph]}}.

\bibitem{Fujimoto:2017lln}
Y.~Fujimoto, T.~Miura, K.~Nishiwaki, and M.~Sakamoto, ``{Dynamical Generation
  of Fermion Mass Hierarchy in an Extra Dimension},''
\href{http://arxiv.org/abs/1709.05693}{{\ttfamily arXiv:1709.05693 [hep-th]}}.

\bibitem{Abe:2008sx}
H.~Abe, K.-S. Choi, T.~Kobayashi, and H.~Ohki, ``{Three Generation Magnetized
  Orbifold Models},''
  \href{http://dx.doi.org/10.1016/j.nuclphysb.2009.02.002}{{\em Nucl. Phys.}
  {\bfseries B814} (2009) 265--292},
\href{http://arxiv.org/abs/0812.3534}{{\ttfamily arXiv:0812.3534 [hep-th]}}.

\bibitem{Abe:2015yva}
T.-h. Abe, Y.~Fujimoto, T.~Kobayashi, T.~Miura, K.~Nishiwaki, M.~Sakamoto, and
  Y.~Tatsuta, ``{Classification of Three-Generation Models on Magnetized
  Orbifolds},'' \href{http://dx.doi.org/10.1016/j.nuclphysb.2015.03.004}{{\em
  Nucl. Phys.} {\bfseries B894} (2015) 374--406},
\href{http://arxiv.org/abs/1501.02787}{{\ttfamily arXiv:1501.02787 [hep-ph]}}.

\bibitem{Mizoguchi:2014gva}
S.~Mizoguchi, ``{F-theory Family Unification},''
  \href{http://dx.doi.org/10.1007/JHEP07(2014)018}{{\em JHEP} {\bfseries 07}
  (2014) 018},
\href{http://arxiv.org/abs/1403.7066}{{\ttfamily arXiv:1403.7066 [hep-th]}}.

\bibitem{Georgi:1974sy}
H.~Georgi and S.~L. Glashow, ``{Unity of All Elementary Particle Forces},''
\href{http://dx.doi.org/10.1103/PhysRevLett.32.438}{{\em Phys. Rev. Lett.}
  {\bfseries 32} (1974) 438--441}.

\bibitem{Inoue:1977qd}
K.~Inoue, A.~Kakuto, and Y.~Nakano, ``{Unification of the Lepton-Quark World by
  the Gauge Group SU(6)},''
\href{http://dx.doi.org/10.1143/PTP.58.630}{{\em Prog.Theor.Phys.} {\bfseries
  58} (1977) 630}.

\bibitem{Fritzsch:1974nn}
H.~Fritzsch and P.~Minkowski, ``{Unified Interactions of Leptons and
  Hadrons},''
\href{http://dx.doi.org/10.1016/0003-4916(75)90211-0}{{\em Ann. Phys.}
  {\bfseries 93} (1975) 193--266}.

\bibitem{Ida:1980ea}
M.~Ida, Y.~Kayama, and T.~Kitazoe, ``{Inclusion of Generations in SO(14)},''
\href{http://dx.doi.org/10.1143/PTP.64.1745}{{\em Prog. Theor. Phys.}
  {\bfseries 64} (1980) 1745}.

\bibitem{Fujimoto:1981bv}
Y.~Fujimoto, ``{SO(18) Unification},''
\href{http://dx.doi.org/10.1103/PhysRevD.26.3183}{{\em Phys. Rev.} {\bfseries
  D26} (1982) 3183}.

\bibitem{Gursey:1975ki}
F.~Gursey, P.~Ramond, and P.~Sikivie, ``{A Universal Gauge Theory Model Based
  on $E_6$},''
\href{http://dx.doi.org/10.1016/0370-2693(76)90417-2}{{\em Phys. Lett.}
  {\bfseries B60} (1976) 177}.

\bibitem{Kojima:2011ad}
K.~Kojima, K.~Takenaga, and T.~Yamashita, ``{Grand Gauge-Higgs Unification},''
  \href{http://dx.doi.org/10.1103/PhysRevD.84.051701}{{\em Phys. Rev.}
  {\bfseries D84} (2011) 051701},
\href{http://arxiv.org/abs/1103.1234}{{\ttfamily arXiv:1103.1234 [hep-ph]}}.

\bibitem{Kojima:2016fvv}
K.~Kojima, K.~Takenaga, and T.~Yamashita, ``{Gauge Symmetry Breaking Patterns
  in an SU(5) Grand Gauge-Higgs Unification Model},''
  \href{http://dx.doi.org/10.1103/PhysRevD.95.015021}{{\em Phys. Rev.}
  {\bfseries D95} no.~1, (2017) 015021},
\href{http://arxiv.org/abs/1608.05496}{{\ttfamily arXiv:1608.05496 [hep-ph]}}.

\bibitem{Burdman:2002se}
G.~Burdman and Y.~Nomura, ``{Unification of Higgs and Gauge Fields in
  Five-Dimensions},''
  \href{http://dx.doi.org/10.1016/S0550-3213(03)00088-9}{{\em Nucl. Phys.}
  {\bfseries B656} (2003) 3--22},
\href{http://arxiv.org/abs/hep-ph/0210257}{{\ttfamily arXiv:hep-ph/0210257
  [hep-ph]}}.

\bibitem{Lim:2007jv}
C.~Lim and N.~Maru, ``{Towards a Realistic Grand Gauge-Higgs Unification},''
  \href{http://dx.doi.org/10.1016/j.physletb.2007.07.053}{{\em Phys.Lett.}
  {\bfseries B653} (2007) 320--324},
\href{http://arxiv.org/abs/0706.1397}{{\ttfamily arXiv:0706.1397 [hep-ph]}}.

\bibitem{Kim:2002im}
H.~D. Kim and S.~Raby, ``{Unification in 5-D SO(10)},''
  \href{http://dx.doi.org/10.1088/1126-6708/2003/01/056}{{\em JHEP} {\bfseries
  01} (2003) 056},
\href{http://arxiv.org/abs/hep-ph/0212348}{{\ttfamily arXiv:hep-ph/0212348
  [hep-ph]}}.

\bibitem{Fukuyama:2008pw}
T.~Fukuyama and N.~Okada, ``{A Simple SO(10) GUT in Five Dimensions},''
  \href{http://dx.doi.org/10.1103/PhysRevD.78.015005}{{\em Phys. Rev.}
  {\bfseries D78} (2008) 015005},
\href{http://arxiv.org/abs/0803.1758}{{\ttfamily arXiv:0803.1758 [hep-ph]}}.

\bibitem{Hosotani:2015hoa}
Y.~Hosotani and N.~Yamatsu, ``{Gauge-Higgs Grand Unification},''
  \href{http://dx.doi.org/10.1093/ptep/ptv153}{{\em Prog. Theor. Exp. Phys.}
  {\bfseries 2015} (2015) 111B01},
\href{http://arxiv.org/abs/1504.03817}{{\ttfamily arXiv:1504.03817 [hep-ph]}}.

\bibitem{Hosotani:2015wmb}
Y.~Hosotani and N.~Yamatsu, ``{Gauge-Higgs Grand Unification},'' {\em PoS}
  {\bfseries PLANCK2015} (2015) 058,
\href{http://arxiv.org/abs/1511.01674}{{\ttfamily arXiv:1511.01674 [hep-ph]}}.

\bibitem{Yamatsu:2015rge}
N.~Yamatsu, ``{Gauge Coupling Unification in Gauge-Higgs Grand Unification},''
  \href{http://dx.doi.org/10.1093/ptep/ptw023}{{\em Prog. Theor. Exp. Phys.}
  {\bfseries 2016} (2016) 043B02},
\href{http://arxiv.org/abs/1512.05559}{{\ttfamily arXiv:1512.05559 [hep-ph]}}.

\bibitem{Furui:2016owe}
A.~Furui, Y.~Hosotani, and N.~Yamatsu, ``{Toward Realistic Gauge-Higgs Grand
  Unification},'' \href{http://dx.doi.org/10.1093/ptep/ptw116}{{\em Prog.
  Theor. Exp. Phys.} {\bfseries 2016} (2016) 093B01},
\href{http://arxiv.org/abs/1606.07222}{{\ttfamily arXiv:1606.07222 [hep-ph]}}.

\bibitem{Hosotani:2016njs}
Y.~Hosotani, ``{Gauge-Higgs EW and Grand Unification},''
  \href{http://dx.doi.org/10.1142/S0217751X16300313}{{\em Int. J. Mod. Phys.}
  {\bfseries A31} no.~20n21, (2016) 1630031},
\href{http://arxiv.org/abs/1606.08108}{{\ttfamily arXiv:1606.08108 [hep-ph]}}.

\bibitem{Hosotani:2017ghg}
Y.~Hosotani and N.~Yamatsu, ``{Gauge-Higgs Seesaw Mechanism in 6-Dimensional
  Grand Unification},'' \href{http://dx.doi.org/10.1093/ptep/ptx124}{{\em Prog.
  Theor. Exp. Phys.} {\bfseries 2017} no.~9, (2017) 091B01},
\href{http://arxiv.org/abs/1706.03503}{{\ttfamily arXiv:1706.03503 [hep-ph]}}.

\bibitem{Hosotani:2017edv}
Y.~Hosotani and N.~Yamatsu, ``{Electroweak Symmetry Breaking and Mass Spectra
  in Six-Dimensional Gauge-Higgs Grand Unification},''
  \href{http://dx.doi.org/10.1093/ptep/ptx175}{{\em Prog. Theor. Exp. Phys.}
  {\bfseries 2018} no.~2, (2018) 023B05},
\href{http://arxiv.org/abs/1710.04811}{{\ttfamily arXiv:1710.04811 [hep-ph]}}.

\bibitem{Slansky:1981yr}
R.~Slansky, ``{Group Theory for Unified Model Building},''
\href{http://dx.doi.org/10.1016/0370-1573(81)90092-2}{{\em Phys. Rept.}
  {\bfseries 79} (1981) 1--128}.

\bibitem{Yamatsu:2015gut}
N.~Yamatsu, ``{Finite-Dimensional Lie Algebras and Their Representations for
  Unified Model Building},''
\href{http://arxiv.org/abs/1511.08771}{{\ttfamily arXiv:1511.08771 [hep-ph]}}.

\bibitem{Ramond:1979py}
P.~Ramond, ``{The Family Group in Grand Unified Theories},'' in {\em
  {International Symposium on Fundamentals of Quantum Theory and Quantum Field
  Theory}}, pp.~265--280.
\newblock 1979.
\newblock
\href{http://arxiv.org/abs/hep-ph/9809459}{{\ttfamily arXiv:hep-ph/9809459
  [hep-ph]}}.
\newblock

\bibitem{Kawamura:2007cm}
Y.~Kawamura, T.~Kinami, and K.-y. Oda, ``{Orbifold Family Unification},''
  \href{http://dx.doi.org/10.1103/PhysRevD.76.035001}{{\em Phys. Rev.}
  {\bfseries D76} (2007) 035001},
\href{http://arxiv.org/abs/hep-ph/0703195}{{\ttfamily arXiv:hep-ph/0703195}}.

\bibitem{Kawamura:2009gr}
Y.~Kawamura and T.~Miura, ``{Orbifold Family Unification in $SO(2N)$ Gauge
  Theory},'' \href{http://dx.doi.org/10.1103/PhysRevD.81.075011}{{\em Phys.
  Rev.} {\bfseries D81} (2010) 075011},
\href{http://arxiv.org/abs/0912.0776}{{\ttfamily arXiv:0912.0776 [hep-ph]}}.

\bibitem{Goto:2013jma}
Y.~Goto, Y.~Kawamura, and T.~Miura, ``{Orbifold Family Unification on Six
  Dimensions},'' \href{http://dx.doi.org/10.1103/PhysRevD.88.055016}{{\em
  Phys.Rev.} {\bfseries D88} no.~5, (2013) 055016},
\href{http://arxiv.org/abs/1307.2631}{{\ttfamily arXiv:1307.2631}}.

\bibitem{Albright:2016lpi}
C.~H. Albright, R.~P. Feger, and T.~W. Kephart, ``{Unification of Gauge,
  Family, and Flavor Symmetries Illustrated in Gauged $SU(12)$ Models},''
  \href{http://dx.doi.org/10.1103/PhysRevD.93.075032}{{\em Phys. Rev.}
  {\bfseries D93} no.~7, (2016) 075032},
\href{http://arxiv.org/abs/1601.07523}{{\ttfamily arXiv:1601.07523 [hep-ph]}}.

\bibitem{Goto:2017zsx}
Y.~Goto and Y.~Kawamura, ``{Orbifold Family Unification Using Vectorlike
  Representation on Six Dimensions},''
\href{http://arxiv.org/abs/1712.06444}{{\ttfamily arXiv:1712.06444 [hep-ph]}}.

\bibitem{Reig:2017nrz}
M.~Reig, J.~W.~F. Valle, C.~A. Vaquera-Araujo, and F.~Wilczek, ``{A Model of
  Comprehensive Unification},''
  \href{http://dx.doi.org/10.1016/j.physletb.2017.10.038}{{\em Phys. Lett.}
  {\bfseries B774} (2017) 667--670},
\href{http://arxiv.org/abs/1706.03116}{{\ttfamily arXiv:1706.03116 [hep-ph]}}.

\bibitem{Reig:2018ocz}
M.~Reig, J.~W.~F. Valle, and F.~Wilczek, ``{$SO(3)$ Family Symmetry and
  Axions},''
\href{http://arxiv.org/abs/1805.08048}{{\ttfamily arXiv:1805.08048 [hep-ph]}}.

\bibitem{Dynkin:1957ek}
E.~Dynkin, ``{Maximal Subgroups of the Classical Groups},'' {\em Amer. Math.
  Soc. Transl.} {\bfseries 6} (1957) 245.

\bibitem{Dynkin:1957um}
E.~Dynkin, ``{Semisimple Subalgebras of Semisimple Lie Algebras},''
{\em Amer. Math. Soc. Transl.} {\bfseries 6} (1957) 111.

\bibitem{Mckay:1977}
W.~Mckay, J.~Patera, and D.~Sankoff, {\em {The Computation of Branching Rules
  for Representations of Semisimple Lie Algebras}}.
\newblock New York Academic Press, 1977.
\newblock in ``Computers in Nonassociative Rings and Algebras'' edited by
  R.~E.~Beck and B.~Kolman.

\bibitem{McKay:1981}
W.~G. McKay and J.~Patera, {\em Tables of Dimensions, Indices, and Branching
  Rules for Representations of Simple Lie Algebras}.
\newblock Marcel Dekker, Inc., New York, 1981.

\bibitem{Cahn:1985wk}
R.~Cahn, {\em {Semi-Simple Lie Algebras and Their Representations}}.
\newblock Benjamin-Cummings Publishing Company,
1985.
\newblock

\bibitem{Fonseca:2011sy}
R.~M. Fonseca, ``{Calculating the Renormalisation Group Equations of a SUSY
  Model with Susyno},'' \href{http://dx.doi.org/10.1016/j.cpc.2012.05.017}{{\em
  Comput.Phys.Commun.} {\bfseries 183} (2012) 2298--2306},
\href{http://arxiv.org/abs/1106.5016}{{\ttfamily arXiv:1106.5016 [hep-ph]}}.

\bibitem{Feger:2012bs}
R.~Feger and T.~W. Kephart, ``{LieART - A Mathematica Application for Lie
  Algebras and Representation Theory},''
  \href{http://dx.doi.org/10.1016/j.cpc.2014.12.023}{{\em Comput.Phys.Commun.}
  {\bfseries 192} (2015) 166--195},
\href{http://arxiv.org/abs/1206.6379}{{\ttfamily arXiv:1206.6379 [math-ph]}}.

\bibitem{Yamatsu:2017sgu}
N.~Yamatsu, ``{Special Grand Unification},''
  \href{http://dx.doi.org/10.1093/ptep/ptx088}{{\em Prog. Theor. Exp. Phys.}
  {\bfseries 2017} no.~6, (2017) 061B01},
\href{http://arxiv.org/abs/1704.08827}{{\ttfamily arXiv:1704.08827 [hep-ph]}}.

\bibitem{Yamatsu:2017ssg}
N.~Yamatsu, ``{String-Inspired Special Grand Unification},''
  \href{http://dx.doi.org/10.1093/ptep/ptx135}{{\em Prog. Theor. Exp. Phys.}
  {\bfseries 2017} no.~10, (2017) 101B01},
\href{http://arxiv.org/abs/1708.02078}{{\ttfamily arXiv:1708.02078 [hep-ph]}}.

\bibitem{Fonseca:2015aoa}
R.~M. Fonseca, ``{On the Chirality of the SM and the Fermion Content of
  GUTs},'' \href{http://dx.doi.org/10.1016/j.nuclphysb.2015.06.012}{{\em Nucl.
  Phys.} {\bfseries B897} (2015) 757--780},
\href{http://arxiv.org/abs/1504.03695}{{\ttfamily arXiv:1504.03695 [hep-ph]}}.

\bibitem{Randall:1999ee}
L.~Randall and R.~Sundrum, ``{A Large Mass Hierarchy from a Small Extra
  Dimension},'' \href{http://dx.doi.org/10.1103/PhysRevLett.83.3370}{{\em Phys.
  Rev. Lett.} {\bfseries 83} (1999) 3370--3373},
\href{http://arxiv.org/abs/hep-ph/9905221}{{\ttfamily arXiv:hep-ph/9905221}}.

\bibitem{Green1984}
M.~B. Green and J.~H. Schwarz, ``{Anomaly Cancellation in Supersymmetric D=10
  Gauge Theory and Superstring Theory},''
\href{http://dx.doi.org/10.1016/0370-2693(84)91565-X}{{\em Phys. Lett.}
  {\bfseries B149} (1984) 117--122}.

\bibitem{Binetruy:1996uv}
P.~Binetruy and E.~Dudas, ``{Gaugino Condensation and the Anomalous $U(1)$},''
  \href{http://dx.doi.org/10.1016/S0370-2693(96)01305-6}{{\em Phys. Lett.}
  {\bfseries B389} (1996) 503--509},
\href{http://arxiv.org/abs/hep-th/9607172}{{\ttfamily arXiv:hep-th/9607172
  [hep-th]}}.

\bibitem{Kojima:2017qbt}
K.~Kojima, K.~Takenaga, and T.~Yamashita, ``{The Standard Model Gauge Symmetry
  from Higher-Rank Unified Groups in Grand Gauge-Higgs Unification Models},''
  \href{http://dx.doi.org/10.1007/JHEP06(2017)018}{{\em JHEP} {\bfseries 06}
  (2017) 018},
\href{http://arxiv.org/abs/1704.04840}{{\ttfamily arXiv:1704.04840 [hep-ph]}}.

\bibitem{Mirabelli:1997aj}
E.~A. Mirabelli and M.~E. Peskin, ``{Transmission of Supersymmetry Breaking
  from a Four-Dimensional Boundary},''
  \href{http://dx.doi.org/10.1103/PhysRevD.58.065002}{{\em Phys. Rev.}
  {\bfseries D58} (1998) 065002},
\href{http://arxiv.org/abs/hep-th/9712214}{{\ttfamily arXiv:hep-th/9712214
  [hep-th]}}.

\bibitem{Hasegawa:2018jze}
K.~Hasegawa and C.~S. Lim, ``{Majorana Neutrino Masses in the Scenario of
  Gauge-Higgs Unification},'' \href{http://dx.doi.org/10.1093/ptep/pty072}{{\em
  PTEP} {\bfseries 2018} no.~7, (2018) 073B01},
\href{http://arxiv.org/abs/1804.04270}{{\ttfamily arXiv:1804.04270 [hep-ph]}}.

\bibitem{Cabibbo:1963yz}
N.~Cabibbo, ``{Unitary Symmetry and Leptonic Decays},''
\href{http://dx.doi.org/10.1103/PhysRevLett.10.531}{{\em Phys. Rev. Lett.}
  {\bfseries 10} (1963) 531--533}.

\bibitem{Kobayashi:1973fv}
M.~Kobayashi and T.~Maskawa, ``{CP Violation in the Renormalizable Theory of
  Weak Interaction},''
\href{http://dx.doi.org/10.1143/PTP.49.652}{{\em Prog. Theor. Phys.} {\bfseries
  49} (1973) 652--657}.

\bibitem{Maki:1962mu}
Z.~Maki, M.~Nakagawa, and S.~Sakata, ``{Remarks on the Unified Model of
  Elementary Particles},''
\href{http://dx.doi.org/10.1143/PTP.28.870}{{\em Prog. Theor. Phys.} {\bfseries
  28} (1962) 870--880}.

\bibitem{Inoue:2000ia}
K.~Inoue and N.-a. Yamashita, ``{Mass Hierarchy from $SU(1,1)$ Horizontal
  Symmetry},'' \href{http://dx.doi.org/10.1143/PTP.104.677}{{\em Prog. Theor.
  Phys.} {\bfseries 104} (2000) 677--689},
\href{http://arxiv.org/abs/hep-ph/0005178}{{\ttfamily arXiv:hep-ph/0005178}}.

\bibitem{Inoue:2003qi}
K.~Inoue and N.-a. Yamashita, ``{Neutrino Masses and Mixing Matrix from
  $SU(1,1)$ Horizontal Symmetry},''
  \href{http://dx.doi.org/10.1143/PTP.110.1087}{{\em Prog. Theor. Phys.}
  {\bfseries 110} (2004) 1087--1094},
\href{http://arxiv.org/abs/hep-ph/0305297}{{\ttfamily arXiv:hep-ph/0305297}}.

\bibitem{Yamatsu:2007}
K.~Inoue and N.~Yamatsu, ``{Charged Lepton and Down-Type Quark Masses in
  $SU(1,1)$ Model and the Structure of Higgs Sector},''
  \href{http://dx.doi.org/10.1143/PTP.119.775}{{\em Prog. Theor. Phys.}
  {\bfseries 119} (2008) 775--796},
\href{http://arxiv.org/abs/0712.2938}{{\ttfamily arXiv:0712.2938 [hep-ph]}}.

\bibitem{Yamatsu:2008}
K.~Inoue and N.~Yamatsu, ``{Strong CP Problem and the Natural Hierarchy of
  Yukawa Couplings},'' \href{http://dx.doi.org/10.1143/PTP.120.1065}{{\em Prog.
  Theor. Phys.} {\bfseries 120} (2008) 1065--1091},
\href{http://arxiv.org/abs/0806.0213}{{\ttfamily arXiv:0806.0213 [hep-ph]}}.

\bibitem{Yamatsu:2009}
K.~Inoue, H.~Kubo, and N.~Yamatsu, ``{Vacuum Structures of Supersymmetric
  Noncompact Gauge Theory},''
  \href{http://dx.doi.org/10.1016/j.nuclphysb.2010.03.004}{{\em Nucl. Phys.}
  {\bfseries B833} (2010) 108--132},
\href{http://arxiv.org/abs/0909.4670}{{\ttfamily arXiv:0909.4670 [hep-th]}}.

\bibitem{Yamatsu:2012}
N.~Yamatsu, ``{New Mixing Structures of Chiral Generations in a Model with
  Noncompact Horizontal Symmetry},''
  \href{http://dx.doi.org/10.1093/ptep/pts079}{{\em Prog. Theor. Exp. Phys.}
  {\bfseries 2013} (2013) 023B03},
\href{http://arxiv.org/abs/1209.6318}{{\ttfamily arXiv:1209.6318 [hep-ph]}}.

\bibitem{Yamatsu:2013}
N.~Yamatsu, ``{A Supersymmetric Grand Unified Model with Noncompact Horizontal
  Symmetry},'' \href{http://dx.doi.org/10.1093/ptep/ptt100}{{\em Prog. Theor.
  Exp. Phys.} {\bfseries 2013} (2013) 123B01},
\href{http://arxiv.org/abs/1304.5215}{{\ttfamily arXiv:1304.5215 [hep-ph]}}.

\end{thebibliography}\endgroup

\end{document}